\documentclass[aps,prl,reprint,amsmath,amssymb,superscriptaddress]{revtex4-1}

\usepackage{graphicx}
\usepackage{dcolumn}
\usepackage{bm}
\usepackage{xcolor}
\usepackage[english]{babel}
\usepackage[hidelinks]{hyperref}
\usepackage{siunitx}

\usepackage{xcolor}
\definecolor{darkblue}{RGB}{0,0,139}

\hypersetup{
    colorlinks=true,
    linkcolor=darkblue,
    citecolor=darkblue,
    urlcolor=darkblue,
    filecolor=darkblue
}
\begin{document}

\title{High-Power Short-Wavelength Free Electron Laser Frequency Combs}

\author{Chenzhi Xu}
\affiliation{Shanghai Institute of Applied Physics, Chinese Academy of Sciences, Shanghai 201800, China}
\affiliation{University of Chinese Academy of Sciences, Beijing 100049, China}

\author{Bingyang Yan}

\affiliation{Shanghai Advanced Research Institute, Chinese Academy of Sciences, Shanghai 201210, China}

\author{Hanxiang Yang}

\affiliation{Shanghai Advanced Research Institute, Chinese Academy of Sciences, Shanghai 201210, China}

\author{Jiawei Yan}
\affiliation{Deutsches Elektronen-Synchrotron DESY, 22603 Hamburg, Germany}

\author{Kai Li}
\email{likai@sari.ac.cn}
\affiliation{Shanghai Advanced Research Institute, Chinese Academy of Sciences, Shanghai 201210, China}

\author{Haixiao Deng}
\email{denghx@sari.ac.cn}
\affiliation{Shanghai Advanced Research Institute, Chinese Academy of Sciences, Shanghai 201210, China}

\begin{abstract}
High-power frequency combs at short wavelengths would open new opportunities for precision measurements and high-resolution spectroscopy. Here, we propose a free-electron laser frequency comb from uncompressed electron bunches. An optical frequency comb serves as the seed, establishing a common phase reference for microbunching in successive electron bunches and transferring its phase coherence to short-wavelength harmonic radiation. By exploiting interference between distinct laser-induced microbunching pathways, we reduce the sensitivity of the harmonic bunching phase to electron-beam energy fluctuations, a major challenge for maintaining interpulse coherence. Start-to-end simulations demonstrate frequency comb generation in the vacuum-ultraviolet ($148\,\mathrm{nm}$) and soft X-ray regimes, yielding average radiation powers of approximately $1.2\,\mathrm{W}$ and $116\,\mathrm{mW}$, respectively, with approximately $8\,\mu\mathrm{W}$ and $110\,\mathrm{nW}$ concentrated in individual comb teeth. These results establish a promising route toward high-power short-wavelength frequency combs for precision spectroscopy, including searches for the nuclear transition in trapped $^{229}$Th ions.

\end{abstract}

\maketitle

Optical frequency combs provide equally spaced, phase coherent spectral lines that bridge optical and radio frequencies, enabling precision spectroscopy, optical clocks, and coherent excitation of quantum systems~\cite{jones2000carrier,hall2006nobel}.
Extending frequency combs to shorter wavelengths opens new opportunities for probing inner-shell electronic and nuclear transitions~\cite{cingoz2012direct,Pupeza2021}.
A particularly compelling example is the exceptionally low energy ($\sim 8.4$~eV) nuclear transition in $^{229}$Th, with a radiative lifetime on the order of $10^3$~s, offering a promising platform for nuclear optical clocks~\cite{peik2003nuclear,elwell2024laser,tiedau2024laser} and tests of fundamental physics~\cite{peik2021nuclear,Fortier:26}.
Recent vacuum-ultraviolet (VUV) frequency comb spectroscopy has improved the precision of the nuclear transition frequency determination in a solid-state host by more than six orders of magnitude~\cite{zhang2024frequency}, paving the way for solid-state $^{229}$Th nuclear clocks~\cite{de2026thorium,huang2026nuclear}.

Short wavelength frequency combs are commonly generated through high harmonic generation (HHG) driven by infrared (IR) optical combs, often using femtosecond enhancement cavities~\cite{Jones2005,Gohle2005}. Although HHG preserves the phase coherence of the driving comb, its low conversion efficiency severely limits the available VUV power. For example, in recent $^{229}$Th spectroscopy experiments, the VUV comb on target is approximately 0.2~mW average power and 1~nW per comb tooth~\cite{zhang2022tunable,zhang2024frequency}. Increasing the frequency comb power would enable more efficient nuclear excitation, and pave the way for coherent manipulation of nuclear transitions and precision spectroscopy of trapped $^{229}$Th ions.

Free-electron lasers (FELs), which extract energy from relativistic electron beams in magnetic undulators, provide a powerful route to high-brightness, wavelength-tunable radiation~\cite{Pellegrini2016,Huang2021}. Previous studies have demonstrated coherent harmonic generation, phase control, and phase-locked pulses using a single electron bunch ~\cite{Allaria2012,Rebernik2019,Gauthier2015PulseShaping,gauthier2016generation,ni2024tunable,hu2025demonstration}. However, transferring and preserving optical phase coherence across successive, independently generated electron bunches while maintaining high radiation power remains a central challenge in realizing FEL-based frequency combs~\cite{PhysRevSTAB.18.030711}.

\begin{figure*}[t]
\centering
\includegraphics[width=\textwidth]{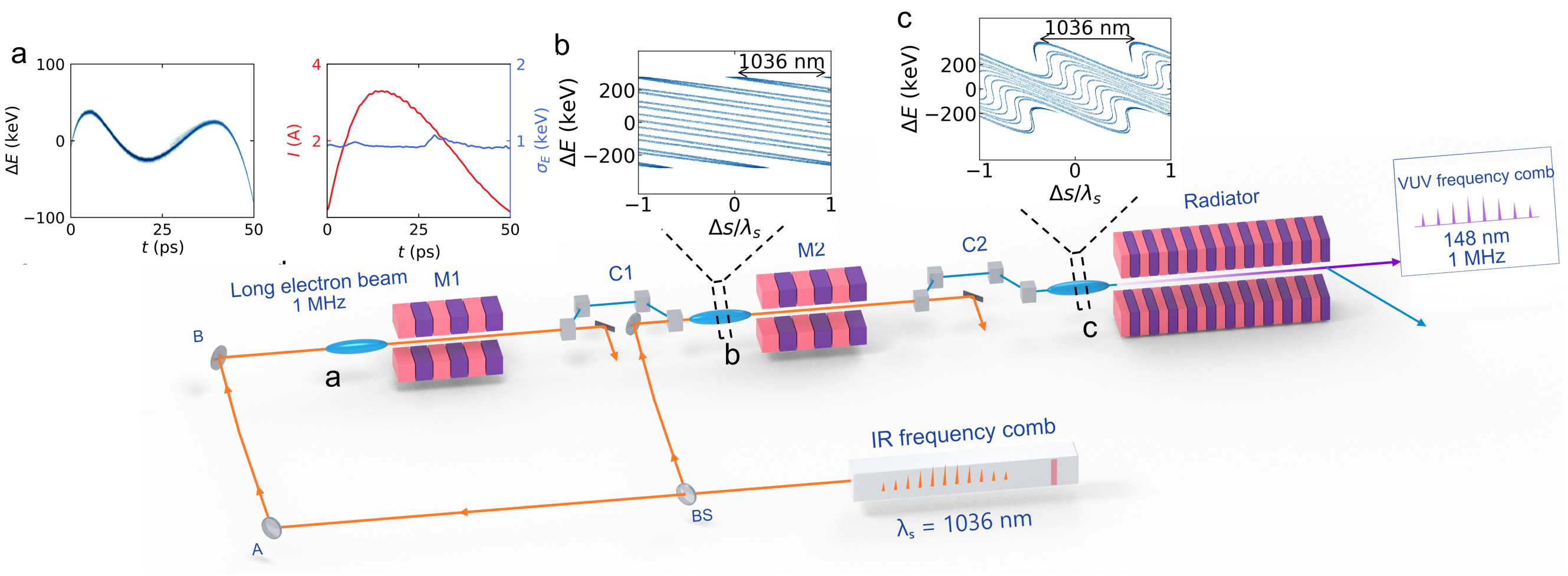}
\caption{Schematic of the proposed high-power, short-wavelength FEL
frequency comb source. Electron bunches at a repetition rate of 1 MHz are seeded by a 1036 nm IR frequency comb. Two successive modulator--chicane stages generate high-harmonic density microbunching through laser-induced
energy modulation and longitudinal phase-space manipulation. The microbunched electron beam subsequently emits coherent radiation at the selected harmonic in a downstream radiator.
(a) Longitudinal phase-space distribution ($\Delta E$), current profile ($I$), and rms slice energy spread ($\sigma_E$) of an uncompressed electron bunch at the entrance.
(b) Enlarged longitudinal phase-space distribution after the first modulator (M1) and chicane (C1), revealing
fine-scale structures formed by energy modulation and dispersive shearing.
(c) Longitudinal phase-space distribution after the second modulator (M2) and chicane (C2), showing the formation of high-harmonic density microbunching.}
\label{fig:scheme}
\end{figure*}

Recent advances in high repetition rate X-ray FEL facilities based on superconducting linear accelerators offer new opportunities for generating frequency combs at megahertz repetition rates. Here, we propose a high-power, short-wavelength FEL frequency comb source based on echo-enabled harmonic generation (EEHG)~\cite{Stupakov2009}, seeded by an infrared optical frequency comb. The optical comb establishes a common phase reference across successive electron bunches, while the EEHG modulation and dispersion parameters are optimized such that the interference between different EEHG microbunching pathways reduces the sensitivity of the harmonic microbunching phase to electron beam central energy fluctuations. We employ long, uncompressed electron bunches to achieve narrow spectral bandwidths and high spectral energy density while minimizing compression-induced phase-space distortions and enhancing radiation phase stability. Using the Shanghai High Repetition Rate XFEL and Extreme Light Facility (SHINE) as a representative platform~\cite{Yan2026AttoSHINE,Chen2026Cryomodules}, start-to-end simulations predict the generation of MHz-spaced frequency combs in the VUV and soft X-ray regimes. 

Figure~\ref{fig:scheme} illustrates the proposed short-wavelength FEL frequency comb scheme. Here, EEHG is employed to upconvert a 1036~nm ytterbium (Yb) optical frequency comb to the seventh harmonic at 148~nm. High-power Yb laser systems can deliver average powers of hundreds of watts~\cite{li2016high,stark20211}, providing a suitable seed source for MHz repetition rate FEL operation~\cite{Yan2021}. The optical pulse train is pulse-picked to 1~MHz to match the electron-bunch repetition rate, stretched, amplified, and injected into the two EEHG modulators. In the first modulator (M1), the seed laser imprints an energy modulation on the electron beam. The first dispersive chicane (C1) then shears the longitudinal phase space, producing fine structures in phase space. The second modulator (M2) introduces an additional energy modulation. The subsequent dispersive chicane (C2) converts the resulting phase-space structures into high-harmonic density microbunching~\cite{Stupakov2009}. The microbunched electron beam subsequently emits coherent radiation in a downstream radiator tuned to resonance at the harmonic wavelength. 

An uncompressed electron bunch with a charge of $100\,\mathrm{pC}$ and a reference energy of $2.2\,\mathrm{GeV}$ is used to demonstrate the proposed scheme. As shown in Fig.~\ref{fig:scheme}(a), it has a peak current of $3.3\,\mathrm{A}$ and a FWHM duration of $30\,\mathrm{ps}$. The accelerator settings for this working point are given in the Supplemental Material~\cite{SupplementalMaterial}. Compared with a strongly compressed bunch, the long, uncompressed bunch has a smaller slice energy spread and energy chirp, while offering higher spectral energy density with a narrower bandwidth. At fixed bunch charge, the one-dimensional FEL saturation power $P_{\mathrm{sat}}\propto I_{\mathrm{pk}}^{4/3}\propto\tau_e^{-4/3}$, where $I_{\mathrm{pk}}$ is the peak current and $\tau_e$ is the electron-bunch duration~\cite{HuangKim2007}. For transform-limited radiation generated in seeded FELs, the spectral bandwidth is $\sim 1/\tau_e$, thus the peak single-pulse spectral energy density $S_{E,\mathrm{peak}}\propto P_{\mathrm{sat}}\tau_e^2\propto\tau_e^{2/3}$. Longer pulses can thus provide higher spectral energy density.

\begin{figure*}[t]
\centering
\includegraphics[width=\textwidth]{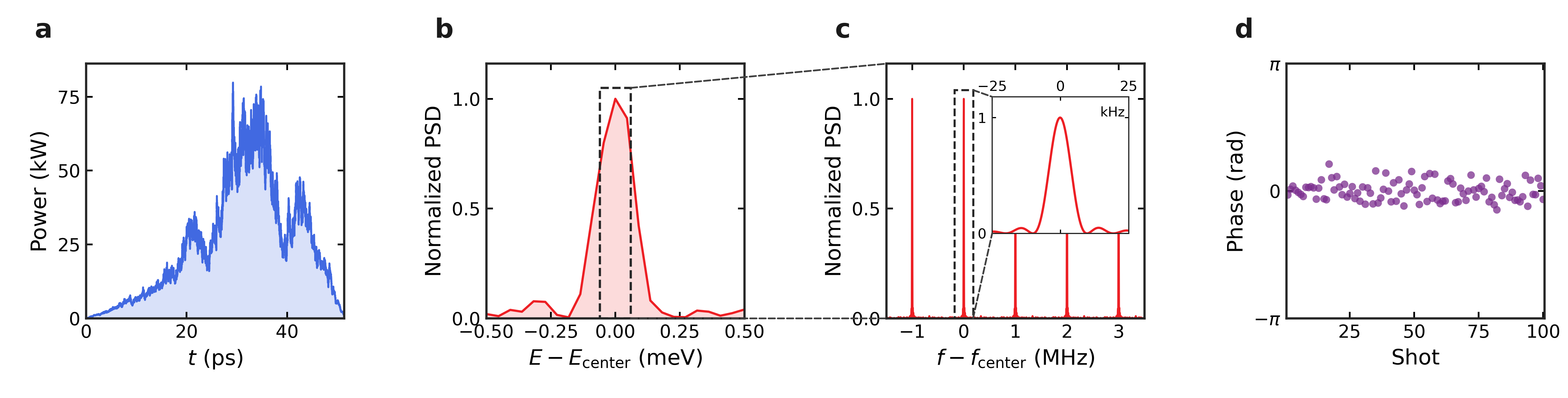}

\caption{Start-to-end simulations of a $148\,\mathrm{nm}$ VUV frequency comb generated by seventh-harmonic EEHG, seeded by a $1036\,\mathrm{nm}$ IR optical frequency comb. (a) Temporal power profile of a representative FEL pulse. (b) Corresponding single-pulse power spectral density (PSD), shown around $E_{\mathrm{center}}$. (c) Frequency comb spectrum obtained from 100 simulated pulses. The comb-tooth spacing is $1\,\mathrm{MHz}$, determined by the pulse repetition rate. The inset shows an individual comb tooth with a linewidth of $8.84\,\mathrm{kHz}$ (FWHM), close to the Fourier limit of $8.86\,\mathrm{kHz}$ for 100 phase-coherent pulses. (d) Phase of each simulated pulse evaluated at $E_{\mathrm{center}}$.}

\label{fig:vuvcomb}
\end{figure*}

Establishing a frequency comb requires phase coherence across radiation pulses. To identify potential phase fluctuations, we trace the phase evolution backward from the emitted radiation to the optical seed. The radiation field inherits the phase of harmonic microbunching~\cite{Gauthier2015PulseShaping,Hemsing2018,Gover2019}, whose bunching factor is written as $\tilde b_h=\bar b_h e^{ih\phi_s}$, where $h$ is the harmonic order and $\phi_s$ is the seed-laser phase. The complex amplitude $\bar b_h$ contains the residual phase introduced by the EEHG process. This phase inheritance enables optical comb coherence to be transferred to short-wavelength radiation. The harmonic microbunching phase in EEHG depends on both the seed laser and the electron beam. Since the phase stability of the optical seed frequency comb can be improved systematically and depends on the specific experimental setup, we assume an ideal optical frequency comb ($\delta\phi_s=0$) and focus on phase variations associated with electron beam centroid-energy and arrival-time fluctuations. The arrival-time fluctuations mainly influence the envelope position instead of the carrier-envelope phase, and thus their effect on interpulse coherence is small. A detailed analysis of the phase fluctuations is provided in the Supplemental Material~\cite{SupplementalMaterial}.

Electron beam centroid-energy fluctuations can introduce phase variations in the harmonic microbunching through energy-dependent longitudinal displacements in the dispersive sections. To illustrate this sensitivity, we consider a single modulator--chicane stage. A centroid-energy offset $\Delta E_c$ produces a longitudinal displacement $\Delta s=R_{56}\Delta E_c/E_0$ and a corresponding harmonic bunching-phase shift $\Delta\Phi_h=k_h\Delta s$, where $k_h$ is the harmonic wavenumber and $E_0$ is the reference electron beam energy. For radiation at $148$~nm, a relative electron beam energy offset of $0.01\%$ and a dispersive strength of $|R_{56}|=1$~mm give a phase shift magnitude of approximately $4$~rad. Such shot-to-shot phase fluctuations can substantially degrade interpulse coherence, posing a challenge for short-wavelength frequency comb generation.

EEHG provides additional degrees of freedom for reducing the sensitivity to centroid-energy fluctuations while retaining harmonic bunching. For two seed lasers with the same wavenumber $k_s$, the harmonic bunching factor in the presence of a normalized centroid-energy offset $q_c=\Delta E_c/\sigma_E$ can be expressed as~\cite{xiang2009echo}
\begin{equation}
\begin{split}
    \bar b_h(q_c)
&=|\bar b_h|e^{i\Phi_h}=\sum_n\tilde C_{n,h}e^{iD_nq_c},\\
D_n&=-(nB_1+hB_2)
\end{split}
\label{eq:eehg_energy_response}
\end{equation}
where $B_i=k_sR_{56,i}\sigma_E/E_0$ is the normalized dispersive strength and $\tilde C_{n,h}$ is the complex amplitude of the $n$th harmonic-generation pathway at the reference energy, which is related to the laser modulation intensity. The phase response of the total bunching depends on the coherent contributions of these pathways. By selecting the modulation amplitudes and dispersive strengths, the sensitivity to centroid-energy fluctuations can be reduced while maintaining useful harmonic bunching. 

\begin{figure*}[t]
\centering
\includegraphics[width=\textwidth]{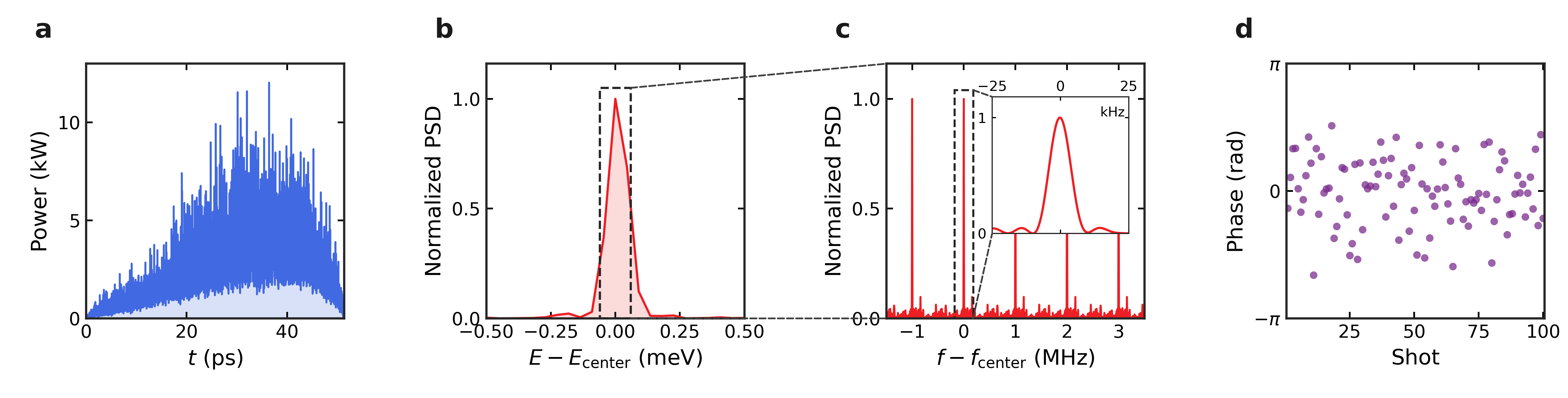}

\caption{Start-to-end simulations of a $10.36\,\mathrm{nm}$ soft X-ray frequency comb generated by 100th-harmonic EEHG, seeded by a $1036\,\mathrm{nm}$ IR optical frequency comb. (a) Temporal power profile of a representative FEL pulse. (b) Corresponding single-pulse PSD shown around $E_{\mathrm{center}}$. (c) Frequency comb PSD. The inset shows an individual comb tooth with a linewidth of $8.96\,\mathrm{kHz}$ (FWHM), close to the finite-train Fourier limit of $8.86\,\mathrm{kHz}$. (d) Phase of each simulated pulse evaluated at $E_{\mathrm{center}}$.}

\label{fig:oxygencomb}
\end{figure*}

We now investigate the proposed scheme using start-to-end simulations. Figure~\ref{fig:vuvcomb} presents the simulated results of a VUV frequency comb generated at the seventh harmonic of a $1036\,\mathrm{nm}$ optical seed, using the FEL code GENESIS 1.3~\cite{reiche1999genesis}. The laser modulation amplitudes and dispersive strengths are optimized to minimize the sensitivity of the microbunching phase to electron-beam centroid-energy fluctuations, $\Delta\Phi_h/\Delta q_c\approx0$, while maintaining an appreciable bunching amplitude $|\bar b_h|$. Each modulator comprises 12 undulator periods of $0.18\,\mathrm{m}$, based on the SHINE configuration. The optimized peak powers of the two seed-laser branches are $9.7\,\mathrm{MW}$ and $1.7\,\mathrm{MW}$, producing peak energy modulations of $\Delta E_1=278\,\mathrm{keV}$ and $\Delta E_2=118\,\mathrm{keV}$, respectively. The chicane dispersive strengths are $R_{56,1}=-27.5\,\mathrm{mm}$ and $R_{56,2}=-4.1\,\mathrm{mm}$. The radiator consists of 14 undulator segments, each with an undulator period of $75\,\mathrm{mm}$ and a magnetic length of $4\,\mathrm{m}$. For the representative pulse shown in Fig.~\ref{fig:vuvcomb}(a), the peak radiation power reaches approximately $75\,\mathrm{kW}$. The corresponding single-pulse spectrum [Fig.~\ref{fig:vuvcomb}(b)] has a full width at half maximum (FWHM) of $0.16\,\mathrm{meV}$. The mean pulse energy is $1.2\,\mu\mathrm{J}$, corresponding to an average radiation power of $1.2\,\mathrm{W}$ at a repetition rate of $1\,\mathrm{MHz}$.

To evaluate interpulse phase coherence and frequency comb formation, we simulate 100 successive electron bunches with accelerator radio-frequency phase and amplitude fluctuations representative of experimental operating conditions. These fluctuations induce variations in the electron-beam centroid energy and arrival time. To preserve the relative phases between successive pulses, the longitudinal coordinate system is referenced to a fixed phase of the optical seed. This reference is maintained throughout the simulations, ensuring that electron-beam longitudinal displacements are retained in the calculated radiation phase. The optimized EEHG and radiator configurations described above are applied to all bunches, yielding the complex radiation fields of $N=100$ successive FEL pulses. The frequency comb spectrum is obtained by coherently summing the complex field spectra without additional phase alignment, $\tilde{E}_{\mathrm{comb}}(f)=\sum_{j=0}^{N-1}\tilde{E}_j(f)e^{-i2\pi f jT}$, where $T=1\,\mu\mathrm{s}$ is the nominal pulse separation and $\tilde{E}_j(f)$ is the complex spectrum of the $j$th pulse. The resulting spectral intensity is proportional to $|\tilde{E}_{\mathrm{comb}}(f)|^2$.

As shown in Fig.~\ref{fig:vuvcomb}(c), the calculated spectrum exhibits well-resolved comb teeth separated by $f_{\mathrm{rep}}=1\,\mathrm{MHz}$. The spectral phase evaluated at the center frequency, $\phi_j=\arg[\tilde{E}_j(f_{\mathrm{center}})]$, has a standard deviation of $\sigma_{\phi}=0.23\,\mathrm{rad}$ [Fig.~\ref{fig:vuvcomb}(d)]. The central comb tooth has a linewidth of $8.84\,\mathrm{kHz}$ (FWHM), in close agreement with the Fourier-limited linewidth of $0.886f_{\mathrm{rep}}/N=8.86\,\mathrm{kHz}$ for a train of 100 phase-coherent pulses. The average power integrated over the central comb tooth, between the frequencies at which the spectral intensity falls to $10\%$ of its peak value, is approximately $8\,\mu\mathrm{W}$. To demonstrate the effects of the phase-sensitivity reduction scheme, the Supplemental Material presents results for EEHG with a normal working point, without phase-sensitivity optimization~\cite{SupplementalMaterial}, which cannot support the formation of a well-resolved frequency comb.

The proposed scheme can be extended to the soft X-ray regime at $10.36\,\mathrm{nm}$, corresponding to the 100th harmonic of the $1036\,\mathrm{nm}$ optical seed. The optimized peak powers of the two seed-laser branches are $9.7\,\mathrm{MW}$ and $9.1\,\mathrm{MW}$, producing peak energy modulations of $\Delta E_1=278\,\mathrm{keV}$ and $\Delta E_2=269\,\mathrm{keV}$, respectively. The chicane dispersive strengths are $R_{56,1}=-140.25\,\mathrm{mm}$ and $R_{56,2}=-1.41\,\mathrm{mm}$, and the radiator comprises 7 undulator segments. A representative pulse is shown in Fig.~\ref{fig:oxygencomb}(a), the peak radiation power reaches approximately $10\,\mathrm{kW}$. The corresponding single-pulse spectrum [Fig.~\ref{fig:oxygencomb}(b)] has a bandwidth of $0.1\,\mathrm{meV}$ (FWHM). The mean pulse energy is $116\,\mathrm{nJ}$, corresponding to an average radiation power of $116\,\mathrm{mW}$ at a repetition rate of $1\,\mathrm{MHz}$.

The frequency comb spectrum is calculated from 100 successive pulses using the coherent summation procedure described above. As shown in Fig.~\ref{fig:oxygencomb}(c), the resulting spectrum exhibits well-resolved comb teeth separated by $1\,\mathrm{MHz}$. The central comb tooth has a linewidth of $8.96\,\mathrm{kHz}$ (FWHM), close to the finite-train Fourier limit of $8.86\,\mathrm{kHz}$ for 100 phase-coherent pulses. The average power integrated over this tooth is approximately $110\,\mathrm{nW}$, using the same integration criterion as for the VUV case. The pulse-to-pulse phase at the center frequency has a standard deviation of $\sigma_{\phi}=0.83\,\mathrm{rad}$ [Fig.~\ref{fig:oxygencomb}(d)].

In summary, we propose a high-power, short-wavelength frequency comb source based on an FEL seeded by an optical frequency comb. Using SHINE as a representative platform, we demonstrate frequency comb generation in the VUV ($148\,\mathrm{nm}$) and soft X-ray ($10.36\,\mathrm{nm}$) regimes. The simulations show that the phase coherence of the optical seed can be transferred to harmonic radiation and preserved across successive electron bunches, enabling the formation of well-resolved comb teeth. The key to maintaining interpulse coherence is the optimization of EEHG parameters to minimize the sensitivity of the harmonic microbunching phase to electron-beam energy fluctuations. Combined with the use of long, uncompressed electron bunches, this approach establishes a promising route toward high-power, phase-coherent frequency combs at short wavelengths for precision metrology and spectroscopy.

A promising application of the proposed VUV frequency comb is precision spectroscopy of the $^{229}$Th nuclear transition. Recent continuous-wave VUV spectroscopy has demonstrated a solid-state nuclear clock~\cite{de2026thorium,huang2026nuclear}, although its precision remains limited by crystal induced inhomogeneous broadening~\cite{ooi2026frequency,li2026origin}. Trapped $^{229}$Th ions offer a complementary platform free from host-crystal perturbations, with the potential for substantially narrower nuclear resonances~\cite{yamaguchi2024laser,beeks2021thorium}. However, locating this nuclear transition precisely remains challenging. A high-power VUV frequency comb could facilitate precision measurements of this transition, providing a route toward improved nuclear clocks.

Experimental verification of interpulse phase coherence represents the most urgent near-term goal. At a repetition rate of $1\,\mathrm{MHz}$, direct interference between successive pulses requires an optical path-length difference of approximately $300\,\mathrm{m}$. As an initial demonstration, frequency comb generation at longer wavelengths, such as $518\,\mathrm{nm}$, could be investigated using a fiber-based delay interferometer. Further near-term studies should also characterize the phase noise of optical-to-FEL phase-transfer process. Looking further ahead, the scheme can be extended to other accelerator-based light sources, particularly in the context of energy recovery linacs~\cite{merminga2003high}. With continued improvements in accelerator stability and electron-beam phase-space control, alternative seeding schemes, such as high-gain harmonic generation and phase-merging enhanced harmonic generation~\cite{Yu1991,PhysRevLett.111.084801}, may also be applied to generate frequency combs.

This work was partially supported by the National Key Research and Development Program of China (2024YFA1612101, 2024YFA1612104), the National Natural Science Foundation of China (12125508, 12541503, 12625515, 12605262), and the International Cooperation Program, Shanghai Municipal Commission of Science and Technology (26140780400).

\bibliographystyle{apsrev4-2}
\bibliography{ssoam}

\end{document}